\documentclass[prb,superscriptaddress,floatfix,showpacs] {revtex4}
\usepackage{amsmath,amssymb,epsfig,psfig,color,pstricks,bm,graphics,alltt}

\newcommand{\BK}{{\ensuremath{\bm{k}}}}
\newcommand{\BG}{{\ensuremath{\bm{G}}}}
\newcommand{\BS}{{\ensuremath{\bm{\Sigma}}}}

\begin{document}

  \title{Evidence for strong electronic correlations in the spectra of
Sr$_2$RuO$_4$}

   \author{Z.V.~Pchelkina}
   \affiliation{Institute of Metal Physics, Russian Academy of
    Sciences-Ural Division, 62041 Yekaterinburg GSP-170, Russia}
   \email{pzv@optics.imp.uran.ru}

   \author{I.A.~Nekrasov}
   \affiliation{Institute of Electrophysics, Russian Academy of
    Sciences-Ural Division, 620016 Yekaterinburg, Russia}

   \author{Th.~Pruschke}
   \affiliation{Institute for Theoretical Physics, University of
    G\"ottingen, Tammannstr.\ 1, 37077 G\"ottingen, Germany}

   \author{A.~Sekiyama}
   \affiliation{Graduate School of
   Engineering Science, Osaka University Toyonaka, Osaka 560-8531 Japan}

   \author{S.~Suga}
   \affiliation{Graduate School of
   Engineering Science, Osaka University Toyonaka, Osaka 560-8531 Japan}

   \author{V.I.~Anisimov}
   \affiliation{Institute of Metal Physics, Russian Academy of
    Sciences-Ural Division, 620219 Yekaterinburg GSP-170, Russia}

   \author{D.~Vollhardt}
   \affiliation{Theoretical Physics III, Center for Electronic
    Correlations and Magnetism, University of Augsburg, 86135
    Augsburg, Germany}

  \date{\today}

  \begin{abstract} The importance of electronic correlation effects in the layered
  perovskite Sr$_2$RuO$_4$ is evidenced. To this end we use state-of-the-art LDA+DMFT (Local Density
  Approximation + Dynamical Mean-Field Theory) in the basis of Wannier functions
  to compute spectral functions and the quasiparticle dispersion of Sr$_2$RuO$_4$.
  The spectra are found to be in good agreement with various spectroscopic
  experiments. We also calculate the $\textbf{k}$-dependence of the quasiparticle bands and
  compare the results with new angle resolved photoemission (ARPES) data. Two typical
  manifestations of strong Coulomb correlations are revealed: (i)  the calculated quasiparticle
  mass enhancement of $m^*/m \approx2.5$ agrees with
  various experimental results, and (ii) the satellite structure at about 3 eV binding energy
  observed in photoemission experiments is shown to be the lower Hubbard
  band. For these reasons Sr$_2$RuO$_4$ is identified as a strongly
  correlated 4$d$ electron material.
  \end{abstract}

  \maketitle

  \section{Introduction}\label{intro}

Intensive research on Sr$_2$RuO$_4$ began after the discovery of
superconductivity at temperatures below 1K.~\cite{Maeno_94} Since it
is widely believed that this system may help to clarify the
mechanism behind high-T$_c$ superconductivity, considerable
theoretical and experimental effort was put into the investigation
of this material which has unconventional properties in spite of a
relatively simple electronic structure. While it is generally
recognized that in 3$d$ transition metal compounds electron
correlations play a crucial role,~\cite{Imada_98} the question
whether the 4$d$ system Sr$_2$RuO$_4$ should be considered a
strongly correlated system, too, or whether its electronic
properties can be understood within conventional band theory
remained open. Indeed, the $4d$ states of the Ru-ion are more
extended than the 3$d$ states and hence correlation effects should
be less significant than, e.g., in high-T$_c$ cuprates. On the other
hand, the effective quasiparticle mass obtained from de Haas-van
Alphen (dHvA),~\cite{Mackenzie_96} ARPES~\cite{arpes_puchkov} and
infrared optical experiments~\cite{optics} is 3-4 times larger than
the results obtained from standard band calculations. 
\cite{Oguchi_95, Singh_95, Hase_96} The temperature-independent
contribution to the magnetic susceptibility and specific heat
constant~\cite{Maeno_94} are also significantly larger than that
given by LDA.~\cite{Oguchi_95} These facts indicate that electron
correlations play an important role in Sr$_2$RuO$_4$.

During the last decade intensive studies of Sr$_2$RuO$_4$ using
various spectroscopic techniques were performed. 
\cite{Schmidt_nexafs,Kurmaev_xas,Okuda_ps,Yokoya_xps,Inoue_96,
Inoue_98,Tran_xps} The evidence derived from these experiments
which point towards correlation effects can be summarized as
follows: (i) the measured bandwidth and density of states (DOS) at
the Fermi level deviate by roughly a factor of 3
\cite{Schmidt_nexafs,Kurmaev_xas} from bandstructure calculations, 
\cite{Oguchi_95, Singh_95, Hase_96}
and (ii) there is a peculiar satellite at $-3$ eV in the
photoemission spectra (PES). \cite{Yokoya_xps,Inoue_96,
Inoue_98,Tran_xps}

The presence of a satellite structure in PES on transition metal
oxides, now commonly interpreted as the lower Hubbard band (LHB), is
generally taken as strong evidence for the importance of
correlations. Such a satellite was first experimentally observed by
Fujimori {\it et al.}~\cite{Fujimori_92} for the $d^1$
perovskite-type Ti$^{3+}$ and V$^{4+}$ oxides. By applying
the \emph{ab initio} LDA+DMFT
approach~\cite{poter97,LDADMFT1,Nekrasov00, psik, LDADMFT, PT} the
corresponding structure in the many-body spectrum was later indeed
identified as the LHB.~\cite{poter97,
Nekrasov00,Zoelfl00,Pavarini_04, sekiyama_sr_ca_vo3}
While in $d^1$ compounds the 3$d$-band is usually well separated
from the oxygen 2$p$ band (making the experimental observation and
theoretical interpretation as the LHB relatively simple) the energy
separation between the Ru-4$d$ and O-2$p$ states  in Sr$_2$RuO$_4$
is much smaller (see below). In this case the LHB may overlap with
the oxygen 2$p$ bands, making the interpretation of structures in
the experimental spectra ambiguous. The difference in experimental
conditions (photon energy, surface sensitivity, sample and surface
quality, etc) complicate the situation even more. Therefore
interpretations of spectroscopic data for Sr$_2$RuO$_4$ are often
controversial.

Early investigations of the electronic structure of Sr$_2$RuO$_4$
used the local density approximation (LDA) to reveal the
similarities and differences with the electronic properties of
cuprate superconductors.~\cite{Oguchi_95,Singh_95} It was then
proposed that the superconductivity of Sr$_2$RuO$_4$ may be
unconventional, namely of triplet type;~\cite{Rice_95, Baskaran_96}
see the comprehensive reviews.~\cite{Mackenzie_03, Bergemann_03} A
quantitative model for triplet superconductivity based on first
principles calculations for the electronic structure and magnetic
susceptibility was suggested by Mazin and Singh.~\cite{Mazin_97} The
electronic structure of Sr$_2$RuO$_4$ and Sr$_2$RhO$_4$ was compared
in~\cite{Hase_96} and the possibility of a magnetic ground state of
Sr$_2$RuO$_4$ was studied within the general gradient approximation
(GGA).~\cite{Boer_99}

The Fermi surface of Sr$_2$RuO$_4$ was also investigated by LDA.
According to these studies, the Fermi surface consists of three
cylindrical sheets, \cite{Oguchi_95, Singh_95, Hase_96, McMullan_96}
in agreement with dHvA experiments.~\cite{Mackenzie_96} By contrast,
ARPES experiments predicted a significantly different Fermi-surface
topology.~\cite{Lu_96, Yokoya_96, Yokoya_961} In principle, such a
discrepancy may be due to strong electronic correlations which are
not taken into account in LDA. However, detailed photoemission
studies~\cite{arpes_puchkov} and scanning-tunneling
microscopy~\cite{Matzdorf_00} subsequently discovered a surface
reconstruction which seemed to resolve the
controversy.~\cite{Damascelli_00,Damascelli_04}

The importance of correlation effects was studied by P\'erez-Navarro
\emph{et al.} \cite{Perez_00} and Arita \emph{et al.}.
\cite{Arita_05}
Although correlations are not very strong, their inclusion was found
to be important for a proper description of the electronic structure
\cite{Perez_00} and a microscopic understanding of superconductivity. 
\cite{Arita_05}

A first multi-band investigation based on the dynamical mean-field
theory (DMFT) to clarify the discrepancy between
dHvA~\cite{Mackenzie_96} and photoemission~\cite{Yokoya_96,
Yokoya_961} data was performed by Liebsch and
Lichtenstein.~\cite{Liebsch_00} They observed a charge flow from the
narrow $xz$, $yz$ bands to the wide $xy$ band leading to a shift of
the van Hove singularity close to E$_F$, and derived quasiparticle
bands by self-consistent second-order perturbation theory for the
self-energy, finding a mass renormalization of 2.1-2.6
~\cite{Liebsch_00} in agreement with experiment.~\cite{arpes_puchkov,
Mackenzie_96} Anisimov {\it et al.} \cite{Anisimov01} investigated
the isoelectric series of alloys Ca$_{2-x}$Sr$_x$RuO$_4$ by means of
LDA+U for $x=0$ and DMFT(NCA) for $0.5<x<2.0$. In the latter doping
range the scenario of an orbital selective Mott transition (OSMT)
was proposed.

In this paper we address the question of the importance of
correlation effects in Sr$_2$RuO$_4$ by a realistic LDA+DMFT(QMC)
calculation within a Wannier function (WF)
formalism.~\cite{wannier_prb} This improved LDA+DMFT(QMC) scheme
allows one to take into account the influence of correlated orbitals
(4$d$-t$_{2g}$ orbitals of Ru in our case) on all other states. This
is essential when $d$ and oxygen $p$ states overlap as is the case
in Sr$_2$RuO$_4$. The comparison between our theoretical results and
experiment clearly shows that electronic correlations have a strong
influence on the electronic structure and lead to the formation of a
pronounced LHB in the spectrum.

The paper is organized as follows. In section \ref{estruc} we
present results for the band structure obtained by LDA (subsection
II.A) and LDA+DMFT(QMC) in the WF basis (subsection II.B),
respectively. Section \ref{exp} contains a comparison of our
LDA+DMFT(QMC) results with XPS (x-ray photoemission spectroscopy),
XAS (x-ray absorption spectroscopy), and NEXAFS (near edge x-ray
fine structure) experiments, as well as with new experimental PES
(photoemission experiments) data (\ref{exp_old}) and recent ARPES
experiments (subsections \ref{exp_XAS} and \ref{exp_arpes}). We
conclude the paper with a summary. \ref{conclusion}

\section{Electronic structure}
\label{estruc}

\subsection{LDA band structure}
\label{lda_band}

\begin{figure}[!h]
\includegraphics[clip=true,width=0.5\textwidth,angle=270]{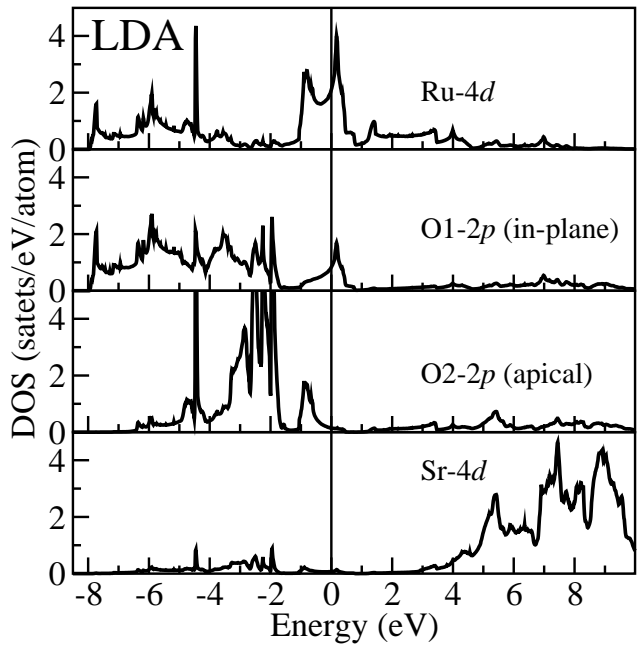}
\caption{\label{lda_dos} Partial LDA DOS for Sr$_2$RuO$_4$. The Fermi level corresponds to zero.}
\end{figure}
\begin{figure}[!h]
\includegraphics[clip=true,width=0.5\textwidth,angle=270]{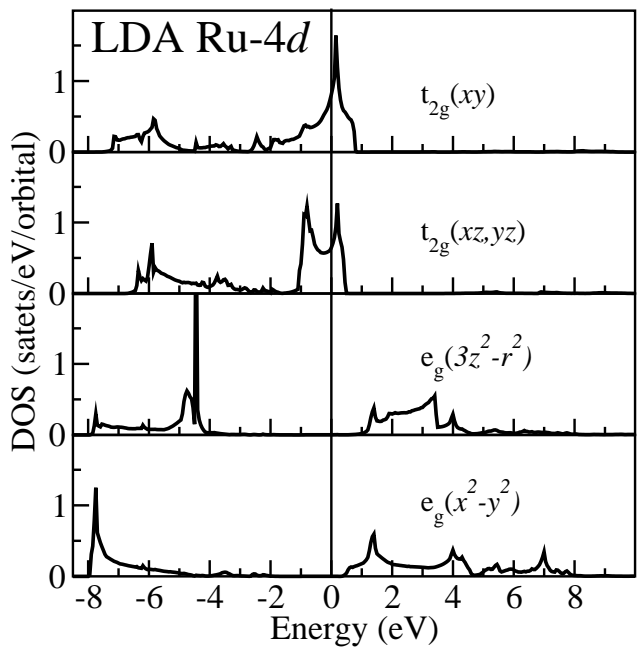}
\caption{\label{Ru_dos} Orbitally projected LDA Ru-4$d$ DOS. The Fermi level corresponds to zero.}
\end{figure}
\begin{figure}[!h]
\includegraphics[clip=true,width=0.3\textwidth,angle=270]{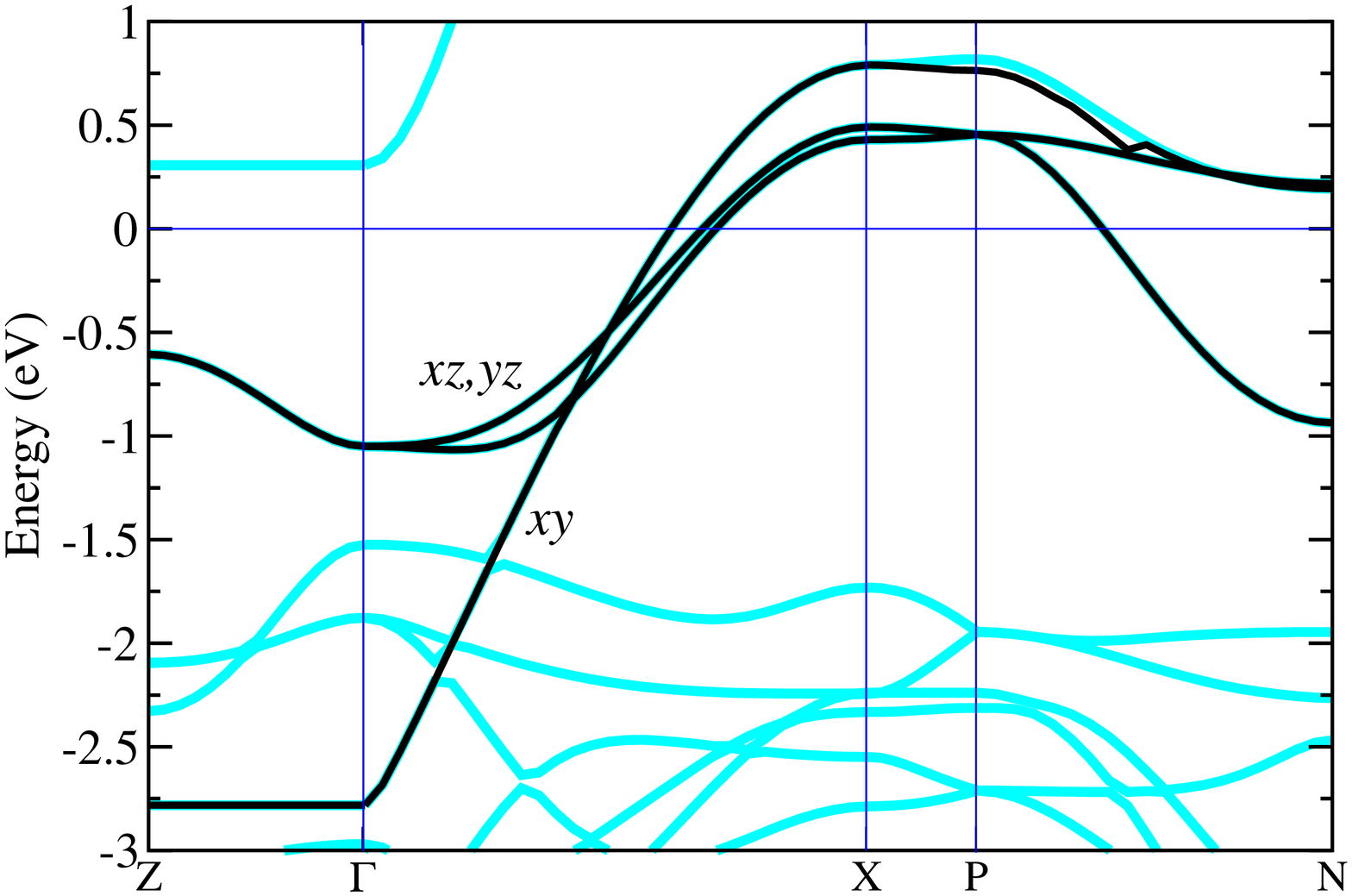}
\caption{\label{lda_bands3} (Colour online) Sr$_2$RuO$_4$ LDA band
structure along high symmetry directions of the Brillouin zone.
Light line - LDA bands, dark line - bands obtained by Wannier
function projection on t$_{2g}$ orbitals. The Fermi level
corresponds to zero.}
\end{figure}

Sr$_2$RuO$_4$ has the undistorted single-layered K$_2$NiF$_4$-type
structure with the space group $I4/mmm$  and lattice parameters
{\emph a}={\emph b}=3.8603~\AA, {\emph c}=12.729~\AA.~\cite{struc}
The structure is formed by layers of RuO$_6$-octahedra separated by
Sr-ions. The RuO$_6$-octahedra are slightly elongated along the
$c$-axis. Therefore the coordination of Ru-ions locally has a
tetragonal symmetry.

Our first-principle calculation of the electronic structure of Sr$_2$RuO$_4$ is based on
density functional theory (DFT) within the LDA approximation~\cite{DFT_LDA, DFT_LDA_1} using
the linearized muffin-tin orbitals (LMTO) method.~\cite{LMTO}
The partial densities of states for Sr$_2$RuO$_4$ are
shown in Fig.~\ref{lda_dos}. They are in good agreement with results of previous
calculations.~\cite{Oguchi_95, Singh_95, Hase_96,  McMullan_96, Boer_99} The strontium
4$d$ states are almost empty and lie above 3 eV; the O-2$p$ derived bands are filled  and
extend from -8 eV to -1 eV.

Physically most interesting are the partially filled ruthenium
4$d$-states. Due to the octahedral coordination of the oxygen ions,
the Ru-4$d$ states are split into t$_{2g}$ and e$_g$ orbitals (see
Fig.~\ref{Ru_dos}). Owing to the stronger hybridization of the two
e$_g$  orbitals with oxygen $p$-states the corresponding bands lie
above the three t$_{2g}$ bands in the energy region from 0.5 eV to 5
eV. In Sr$_2$RuO$_4$ four Ru-4$d$ electrons occupy the three
t$_{2g}$ bands ($d^4$ configuration). The layered crystal structure
of Sr$_2$RuO$_4$ results in a two-dimensional DOS of the $xy$
orbital while the $xz$, $yz$ orbitals have nearly one-dimensional
character (see Fig.~\ref{Ru_dos}). The $xy$ orbital hybridizes with
$xy$ orbitals of the four Ru neighbors and thus has a bandwidth
almost twice as large (W$_{xy}$=2.8 eV) as that of the $xz$, $yz$
orbitals (W$_{xz,yz}$=1.5 eV) which hybridize with corresponding
orbitals of two Ru neighbors only.

The LDA bands in the energy window $-3\ldots1$ eV are shown in
Fig.~\ref{lda_bands3}. In contrast to a typical $d^1$
systems,~\cite{sekiyama_sr_ca_vo3} there is no well-pronounced
separation of the oxygen 2$p$ and ruthenium 4$d$ states in
Sr$_2$RuO$_4$. More
precisely, Fig.~\ref{lda_bands3} shows that while the Ru-4$d$ $xz$,
$yz$ orbitals are separated from the oxygen 2$p$ bands the Ru-4$d$
$xy$ orbital strongly overlaps with these oxygen bands.

\subsection{LDA+DMFT(QMC) results: effect of correlations}
\label{corr_band}

Sr$_2$RuO$_4$ is a paramagnetic metal. \cite{Maeno_94}
It is well known that the paramagnetic state of a correlated metal
is well described by the DMFT (for reviews see
\cite{vollha93, pruschke, georges96}). Within DMFT the electronic
lattice problem is mapped onto a single-impurity Anderson model with
a self-consistency condition.~\cite{Georges92,Jarrell92}
This mapping, which becomes exact in the limit of large coordination
number of the lattice,~\cite{DMFT_vollha} allows one to investigate
the dynamics of correlated lattice electrons non-perturbatively at
all interaction strengths. We use the LDA+DMFT {\em ab initio}
technique~\cite{poter97,LDADMFT1,Nekrasov00, Zoelfl00} (for
an introduction see, \cite{PT} for reviews see \cite{psik,
LDADMFT,Kotliar_review,Held05})
to investigate correlation effects in Sr$_2$RuO$_4$. The effective
impurity problem corresponding to the many-body Hamiltonian is
solved by quantum Monte Carlo simulations.~\cite{QMC} The LDA+DMFT
approach was recently improved by employing a Wannier functions (WF)
formalism,~\cite{wannier_prb} which allows one to project the
Hamilton matrix from the full-orbital space to a selected set of
relevant orbitals. The projection ensures that the information about
all states in the system is kept. In the present work we use the WF
formalism to construct an effective few-orbital Hamiltonian with
t$_{2g}$ symmetry and to take into account the influence of
correlated t$_{2g}$-orbitals on other states. A three-orbital
Hamiltonian obtained by the WF projection with dispersions presented
in Fig.~\ref{lda_bands3} (black lines) was used as an {\it ab
initio} setup of the correlation problem. {\em Ab initio} values of
the orbitally averaged Coulomb interaction parameter $\bar U$=1.7 eV
and Hund exchange parameter $J$=0.7 eV were obtained by
constrained LDA calculations.~\cite{Gunnarson_89} We emphasize that
not only the on-site e$_g$ screening and the screening from Ru-ions
of the RuO$_2$ plane were taken into account in the calculation of
the Coulomb interaction parameter but also screening from
neighboring RuO$_2$ planes.

In the particular case of three t$_{2g}$-orbitals $\bar U$ is equal
to the inter-orbital Coulomb repulsion $U'$.~\cite{poter97, LDADMFT1} Thus we
obtain $U=U'+2J=$3.1 eV for the local intra-orbital Coulomb
repulsion. We note that our values of $U$ and $J$ differ
substantially from those by Liebsch and
Lichtenstein~\cite{Liebsch_00} who assumed a much smaller Hund
exchange parameter. These authors estimated the Coulomb repulsion
parameter from the XPS spectrum~\cite{Yokoya_96} using the position
of the resonance satellite. The value of about 1.5 eV obtained
thereby agrees well with our calculated value for $\bar U$.

\begin{figure}[!h]
\includegraphics[clip=true,width=0.35\textwidth,angle=270]{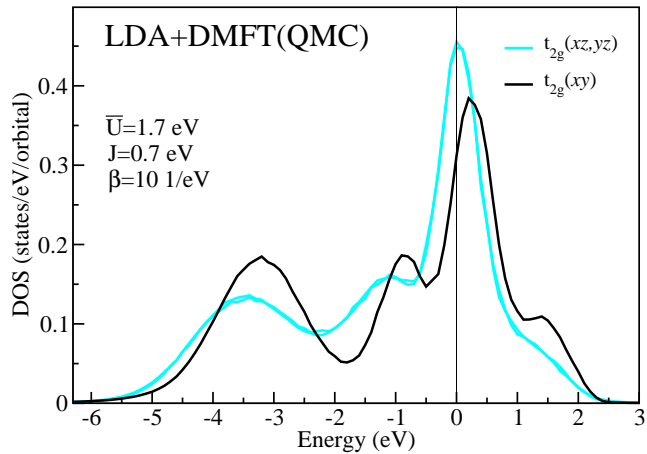}
\caption{\label{qmc_dos} (Colour online) Ru-4$d$(t$_{2g}$) spectral
functions obtained within LDA+DMFT (QMC) using a projected
Hamiltonian. Dark curve:  $xy$ orbital, light curve: $xz$, $yz$
orbitals; $\bar U$=1.7 eV, $J$=0.7 eV. The Fermi level corresponds
to zero.}
\end{figure}

\begin{figure}[!h]
\includegraphics[clip=true,width=0.35\textwidth,angle=270]{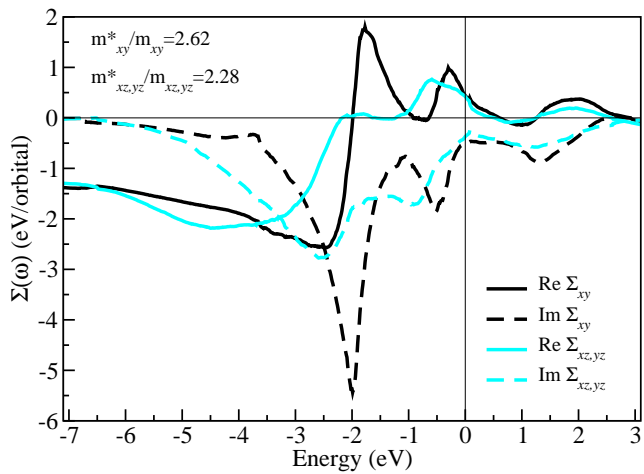}
\caption{\label{sigma} (Colour online) Self-energy on the real
energy axis for $xy$ (black line) and $xz$, $yz$ (light line)
orbitals of Sr$_2$RuO$_4$. Solid line: real part; dashed line:
imaginary part. The Fermi level corresponds to zero.}
\end{figure}

The three-orbital, projected Hamiltonian together with the {\it ab
initio} Coulomb interaction parameters were used as input for the
QMC simulation of the effective quantum impurity problem arising in
the DMFT. The simulations were performed for an inverse temperature
$\beta$=10 eV$^{-1}$ using 40 imaginary time slices ($\Delta
\tau$=0.25). The imaginary time QMC data were analytically continued
by maximum entropy.~\cite{MEM} The results are shown in
Fig.~\ref{qmc_dos}. We find a pronounced lower Hubbard band (LHB)
between -5 and -1 eV,  a quasiparticle peak (QP) around the Fermi
level, and an upper Hubbard band (UHB) at about 1.5 eV. Real and
imaginary parts of the corresponding self-energy for real
frequencies (for details see Appendix B in \cite{wannier_prb}) for
t$_{2g}$ orbitals are shown in Fig.~\ref{sigma}. The mass
enhancement calculated from the derivative of Re$\Sigma$ at the
Fermi level amounts to 2.62 for the $xy$ orbital  and 2.28 for the
$xz$, $yz$ orbitals, in agreement with results from ARPES, 
\cite{arpes_puchkov} dHvA~\cite{Mackenzie_96} and infrared optical
experiments. \cite{optics} A detailed analysis of the structures
seen in Fig.~\ref{qmc_dos} is presented in Appendix \ref{thomas}.

\begin{figure}[!h]
\includegraphics[clip=true,width=0.55\textwidth,angle=270]{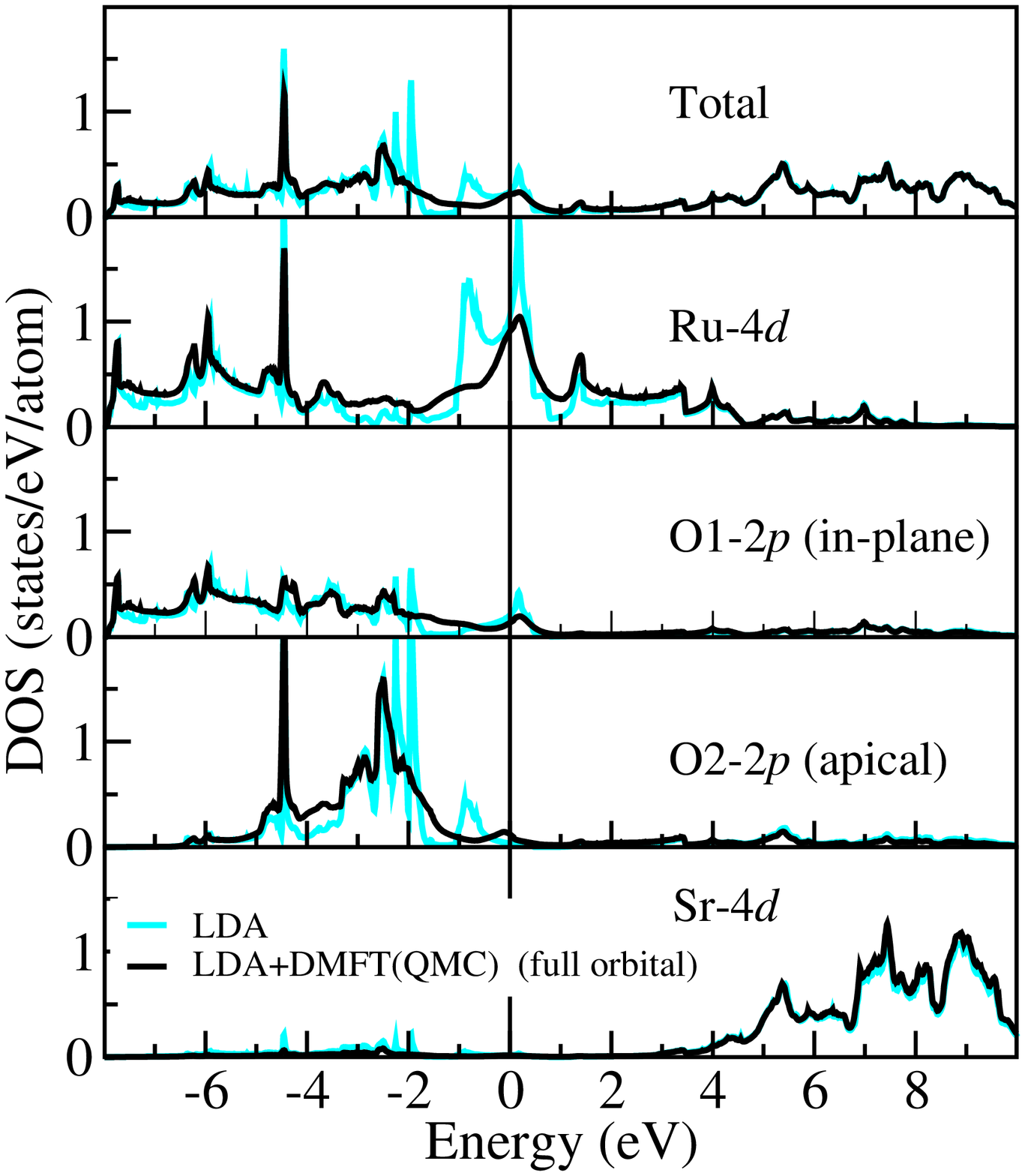}
\caption{\label{bt_dos} (Colour online) Comparison of the total and
partial LDA DOS (light curve) and the DOS calculated using the
full-orbital self-energy from LDA+DMFT(QMC) (black curve). The Fermi
level corresponds to zero.}
\end{figure}

After having calculated the self-energy $\Sigma$($\omega$) for real
frequencies one may perform the inverse transformation from the
reduced Wannier basis back to the full LMTO basis. 
\cite{wannier_prb} This step allows one to take into account the
influence of the three correlated t$_{2g}$ orbitals on all other
states of ruthenium, oxygen and strontium. The comparison of the
noninteracting LDA partial density of states with the one obtained
by the inverse transformation is shown in Fig.~\ref{bt_dos}. Since
the hybridization of Ru and O is quite strong the oxygen states are
changed rather significantly by correlation effects on Ru-ions.
These changes are most pronounced in the energy region between -4
and -1 eV. One can see that apical oxygen atoms are more affected by
correlations than in-plane atoms. We believe that this is due to the
one-dimensional character of the $xz,yz$ states of the Ru-4$d$
shell. Hence correlation effects are much stronger for these
orbitals; consequently the DOS for the apical oxygen atoms is
strongly modified. The strontium states are less affected.

\begin{figure}[!h]
\includegraphics[clip=true,width=0.55\textwidth,angle=270]{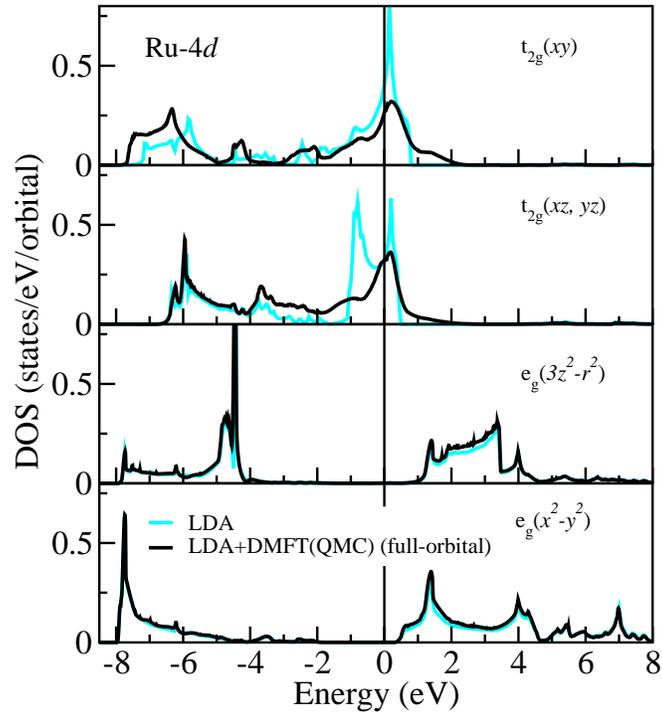}
\caption{\label{bt_Ru_dos} (Colour online) The same as in Fig.~\ref{bt_dos} but
for Ru-4$d$ states only.}
\end{figure}

A comparison between the partial LDA DOS of Ru-4$d$  and the DOS
obtained using the full-orbital self-energy from our LDA+DMFT(QMC)
calculations is shown in Fig.~\ref{bt_Ru_dos}. The main effect of
the correlations on the Ru site is seen to be a transfer of spectral
weight from the energy region near the Fermi level to the lower and
upper Hubbard bands range from -4 eV to -1 eV, and from 1 eV to 2 eV
respectively.

\section{Comparison with experimental data}
\label{exp}
\subsection{XPS and PES experiments}
\label{exp_old} We will now compare the computed LDA+DMFT(QMC)
spectral functions for the $t_{2g}$ electrons (light solid lines in
Fig.~\ref{pes_xps}, \ref{suga_pes}, \ref{kurmaev_xas}) and those
calculated using the full-orbital self-energy (black solid lines in
the same figures) with several experimentally obtained spectra
describing both valence and conduction bands.
To compare with experiment we took into account the photoemission
cross section ratio for Ru-4$d$ and O-2$p$ states as a function of
photon energy.~\cite{cross_sec} We found that, in general, an energy
dependent broadening of the theoretical spectral functions gives
better agreement with the experimental data (see~\cite{Liu_92} and
\cite{ Arko_00}). For this the theory curves were convoluted using a
Gaussian with a full width at half maximum increasing as $C
\cdot$E+$g$. Here $E$ is the binding energy, $g$ is the experimental
resolution, and $C$ characterizes the increase of the broadening
with energy upon moving away from Fermi level due to core-hole life
time effects. The maximally allowed broadening was restricted to 1~
eV. Specific values of $C$ and $g$ parameters used for comparison
with experiment are indicated in the corresponding figures.

\subsubsection{Comparison with previous XPS experiments}
\begin{figure*}[!h]
\includegraphics[clip=true,width=0.35\textwidth,angle=270]{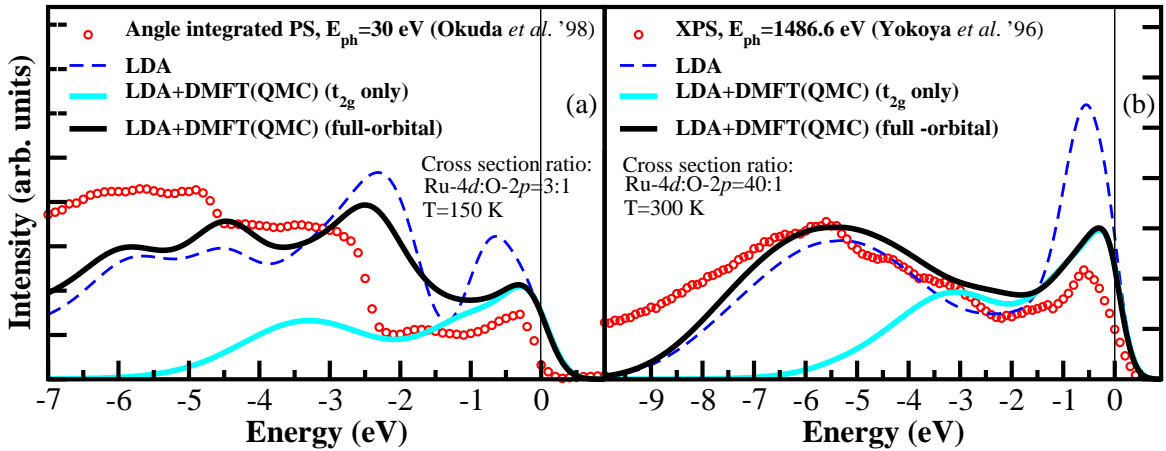}
\caption{\label{pes_xps} (Colour online) Theoretical spectral
functions of Sr$_2$RuO$_4$ calculated by LDA+DMFT(QMC); light solid
line: t$_{2g}$ orbitals, black solid line: full-orbital self-energy.
In (a) the results are compared to an angle-integrated valence band
photoemission spectrum~\cite{Okuda_ps} obtained with a photon energy
$E_{\rm ph}=30$ eV, while in (b) we compare to an XPS spectrum
obtained with a photon energy $E_{\rm ph}=1486.6$
eV.~\cite{Yokoya_xps} The theoretical spectra are convoluted using a
linear broadening -0.04$\cdot$E+0.25 for (a), and -0.14$\cdot$E+0.25
for (b) to account for the experimental resolution. Intensities are
normalized on the area under curves. The Fermi level corresponds to
zero.}
\end{figure*}

In Fig.~\ref{pes_xps} we compare an angle integrated valence band
photoemission spectrum  of Sr$_2$RuO$_4$~\cite{Okuda_ps} 
[Fig.~\ref{pes_xps}(a)] and the XPS spectra from
Ref.~\cite{Yokoya_xps} [Fig.~\ref{pes_xps}(b)] with the theoretical
spectral functions. The contributions from Ru-4$d$ and O-2$p$
spectra were weighted according to the photoemission cross section
ratio \cite{cross_sec} 3:1 for Fig.~\ref{pes_xps}(a), corresponding to
a photon energy 30 eV, and 40:1 for Fig.~\ref{pes_xps}(b),
corresponding to a photon energy of 1486.6 eV. The theoretical
spectra were multiplied with the Fermi function at $T=150$ K and
$300$ K, respectively. In Fig.~\ref{pes_xps} a linear broadening was
employed. UPS (ultra-violet photoemission) data of the valence band
of Sr$_2$RuO$_4$~\cite{Schmidt_nexafs} obtained  at higher photon
energy of 60~eV and 110~eV show similar features as the PES and XPS
spectra in Fig.~\ref{pes_xps}.

We note that for energies below -2 eV  the weight of oxygen states
in the spectrum (according to the cross section ratio) becomes
essential. Indeed, only by proper inclusion of this contribution can
a satisfactory description of the experimental spectra with the
full-orbital LDA+DMFT(QMC) spectral function be achieved.

The experimental spectrum in Fig.~\ref{pes_xps}(a) was obtained at a
rather low photon energy. Therefore, according to the cross section
ratio, the contribution of oxygen states is considerable. The
calculated curves reproduce all features of the complicated
structure of the experimental data, although the positions and
weight of peaks agree only qualitatively. The most serious
disagreement with experiment can be observed in the energy region
from -2.5 eV up to the Fermi level. Despite similar line shapes the
theoretical curves have too little weight. This failure can, for
example, be attributed to an underestimation of the oxygen
contribution in the theoretical curves, or to matrix element effects
which may be significant at low photon energies. Nevertheless, the
LDA+DMFT(QMC) spectra are seen to be in much better agreement with
experiment than the LDA results. In particular, in LDA+DMFT the LDA
peak near -0.5 eV is merely becomes a plateau -- the redistribution
of spectral weight obviously being an effect of correlations.
Comparing the theoretical t$_{2g}$ and full-orbital spectra with
experiment, one can see that the latter yield a much better
description in a wide energy range.

The XPS spectrum in Fig.~\ref{pes_xps}(b) obtained at a very high
photon energy is seen to be almost exclusively Ru-4$d$
states. Obviously, the full--orbital spectral function gives good  agreement
with the XPS data. Moreover, one observes a pronounced maximum at -3
eV which experimentalists previously interpreted as the
LHB.~\cite{Yokoya_xps} This conjecture is now confirmed by our
calculations (see the detailed discussion below).

\subsubsection{Comparison with new PES experiment}
\begin{figure}[!h]
\includegraphics[clip=true,width=0.35\textwidth,angle=270]{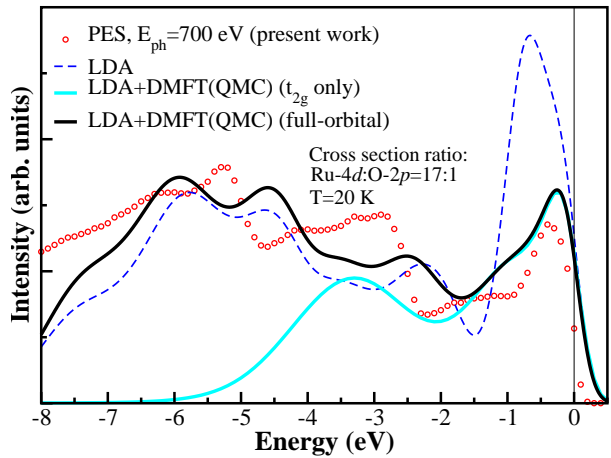}
\caption{\label{suga_pes} (Colour online) Similar plot as in
Fig.~\ref{pes_xps} but now we compare with PES spectrum of
Sr$_2$RuO$_4$. The theoretical spectra are convoluted using a linear
broadening -0.04$\cdot$E+0.20 to account for the experimental
resolution. }
\end{figure}

Clean (001)  surfaces of high purity single crystal samples were
obtained by \emph{in situ} cleavage  at 20 K in ultrahigh vacuum.
Angle-integrated and angle-resolved spectra were measured using the
GAMMADATA-SCIENTA SES200 Analyzer at BL25SU of SPring-8 by use of
circularly polarized light. Both measurements were performed at 700
eV by detecting near normal emission electrons to obtain highest
bulk sensitivity at this photon energy. The resolution for these
measurements was set to 200 meV (Fig.~\ref{suga_pes}) and 120 meV
(Fig.~\ref{dispers_comp}), respectively. The energy scale was
calibrated by the Fermi edge of Au. The surface cleanliness was
checked by the absence of possible additional spectral weight on the
higher binding energy side of the O 1$s$ peak, and by the absence of
the C 1$s$ contribution.

In Fig.~\ref{suga_pes} we compare photoemission spectrum of
Sr$_2$RuO$_4$ with spectral functions calculated by LDA+DMFT(QMC). A
weighted sum of Ru-4$d$ and O-2$p$ spectral functions according to
the photoemission cross section ratio 17:1~\cite{cross_sec} was used
-- corresponding to an experimental photon energy of 700~eV.
Theoretical spectra were multiplied with the Fermi function at 20 K
and were linearly broadened to account for the experimental
resolution.

\subsubsection{Interpretation} 
We now discuss and interpret the experimental and theoretical
spectra, and also check the presence of a LHB in the computed
spectra. In Fig.~\ref{pes_xps}(b) the structure in the experimental
XPS-spectrum at -3 eV  was interpreted as the LHB. \cite{Yokoya_xps}
In the full-orbital LDA+DMFT(QMC) spectral function (black solid
line) a corresponding feature is indeed visible, but has less
intensity and appears only as a shallow shoulder rather than a
distinct bulge. This structure is obviously not described
by LDA (dashed line). Looking at the light solid line in
Fig.~\ref{pes_xps}(b) which represents the LDA+DMFT(QMC) DOS for the
t$_{2g}$ orbitals alone, we are able to identify this shoulder
unambiguously as a result of the LHB. Thus we have theoretically
confirmed the interpretation of Yokoya \emph{et
al.}.~\cite{Yokoya_xps} A very similar feature corresponding to the
LHB in our LDA+DMFT(QMC) DOS was reported in Ref.~\cite{Tran_xps}.

The situation is similar in the PES spectrum. Because of lower
photon energies, and due to the enhancement of the O-2$p$
contribution in the PES spectrum seen in Fig.~\ref{suga_pes}, one
again cannot identify the -3 eV satellite directly. However, one can
recognize a feature in the spectrum whose position coincides with
the LHB in our LDA+DMFT(QMC) calculations. In the first theoretical
DMFT work on the ruthenate,~\cite{Liebsch_00} model-Coulomb
parameters were chosen as $\bar U$=0.8 eV and $J$=0.2 eV. As a
consequence a less well-defined LHB with low spectral weight was
obtained.

\subsection{XAS and NEXAFS experiments}
\label{exp_XAS}

\begin{figure*}[!h]
\includegraphics[clip=true,width=0.35\textwidth,angle=270]{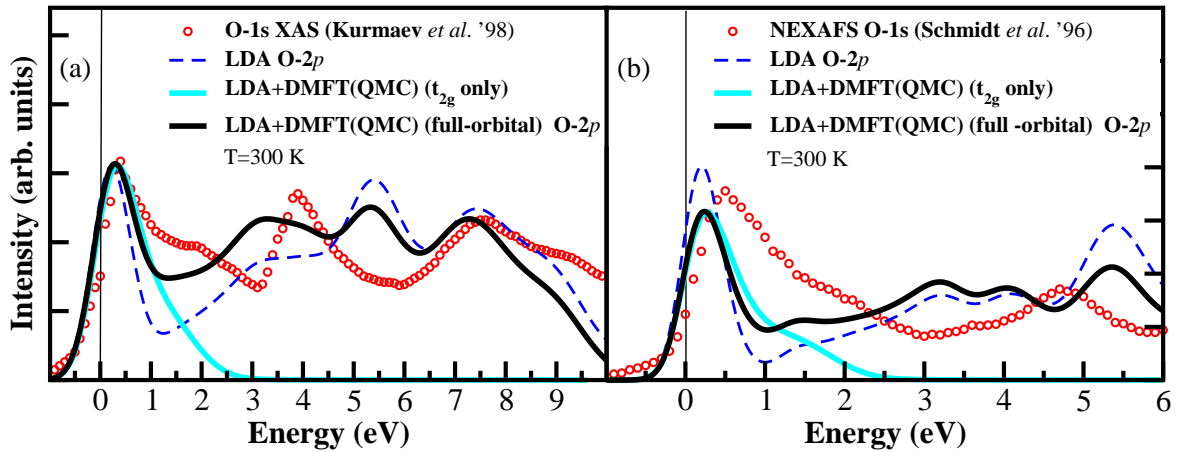}
\caption{\label{kurmaev_xas} (Colour online) Theoretical spectra of
Sr$_2$RuO$_4$ calculated by LDA+DMFT(QMC); light solid line: only
t$_{2g}$ electrons,  black solid line: full-orbital self-energy
using O-2p states. (a) Comparison with O-1$s$ XAS
spectrum,~\cite{Kurmaev_xas} (b) comparison with NEXAFS
spectrum.~\cite{Schmidt_nexafs} Theoretical spectra are convoluted
using linear broadening 0.01$\cdot$E+0.35 for (a), and
0.02$\cdot$E+0.2 for (b) to account for the experimental resolution.
Intensities are normalized on the area under curves. The Fermi level
corresponds to zero.}
\end{figure*}

In Fig.~\ref{kurmaev_xas}(a) the O-1$s$ x-ray absorption spectrum
(XAS) of Sr$_2$RuO$_4$~\cite{Kurmaev_xas} representing the
conduction band is compared with the O-2$p$ spectral function
calculated via LDA+DMFT(QMC). Furthermore, Fig.~\ref{kurmaev_xas}(b)
shows the near edge x-ray fine structure spectrum (NEXAFS) of
Sr$_2$RuO$_4$~\cite{Schmidt_nexafs} together with the theoretical
spectral functions. The theoretical spectra are multiplied with the
Fermi function at $T=300$ K and convoluted using linear broadening
to account for the experimental resolution.

The agreement between theory and experiment in
Fig.~\ref{kurmaev_xas} is found to be only qualitative. This may be
due to empty states (conduction band) in the LMTO method. Namely,
the conventional LMTO choice of the MTO linearization energy point
lies inside the occupied part of the bands. As a result the
unoccupied states in LDA calculation are not properly described.
Nevertheless, due to the spectral weight redistribution in DMFT
calculation, the agreement with experiment is improved (black solid
line) in comparison with the bare LDA O-2$p$ DOS (light dashed
line).

\subsection{High-energy bulk ARPES experiment}
\label{exp_arpes}

The comparison of experimental and theoretical spectra presented in
the previous sections proves the existence of a LHB in
Sr$_2$RuO$_4$. Hence Sr$_2$RuO$_4$ must be considered a strongly
correlated electron material. Further evidence for correlation
effects comes from the renormalization of quasiparticle properties,
i.e., effective masses and band dispersions. This
renormalization was addressed by Liebsch and
Lichtenstein~\cite{Liebsch_00} who derived quasiparticle bands from
self-consistent second-order self-energy. The Coulomb repulsion and
Hund's rule exchange parameters chosen by these authors were rather
small ($U$=1.2-1.5 eV, $J$=0.2-0.4 eV), resulting in a LHB with low
spectral weight; nevertheless the effective masses
($m^{\ast}\approx2.1-2.6$) were found to be in good agreement with
experiments. In our investigation we determined the local Coulomb
repulsion by the {\em ab initio} constrained LDA yielding $\bar
U$=1.7 eV and $J$=0.7 eV, i.e., a value of $U$=3.1 eV which is twice as large as
that used in Ref.~\cite{Liebsch_00}. This value of $U$ now produces
a pronounced LHB and, at the same time, gives almost the same values
of $m^{\ast}/m$ as those reported in Ref. ~\cite{Liebsch_00}:
$m^{\ast}/m$=2.62 for the $xy$ orbital and 2.28 for the $xz$, $yz$
orbitals, respectively. All these values are in good agreement with
experimental estimations. \cite{arpes_puchkov, Mackenzie_96, optics}

\begin{figure}[!h]
\includegraphics[clip=true,width=0.7\textwidth]{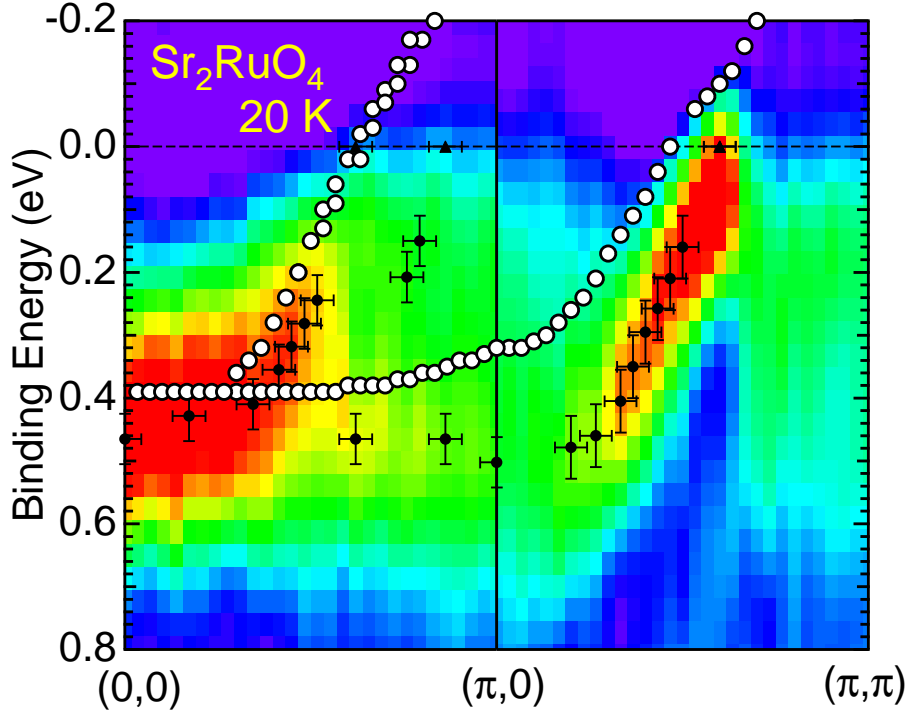}
\caption{(Colour online) Comparison of the dispersion extracted from
high-energy ARPES data~\cite{Suga_ARPES} and LDA+DMFT(QMC).
Experimental data contain the second derivative of the energy
distribution curves, peak positions of the second derivative
(black dots), and k$_F$ estimated from the momentum
distribution curves at E$_F$ (triangles). The theoretical dispersion
is indicated by open circles.}\label{dispers_comp}
\end{figure}

We will now proceed to calculate \emph{${\BK}$-resolved} spectra and
quasiparticle band dispersions following the strategy proposed in
Ref.~\cite{Liebsch_00, kink_paper, Biermann01} and employed
by us to calculate the ARPES spectra of SrVO$_3$. \cite{kink_paper}
ARPES computations have previously been performed also for the 2D
Hubbard model by Maier~{\em et al.}~\cite{Maier02} in the framework
of the dynamical cluster approximation (DCA),~\cite{dca} and by
Sadovskii~{\em et al.}~\cite{Sadovskii05} within the so-called
DMFT+$\Sigma_{\BK}$ approach.

To compare theoretical quasiparticle bands with the dispersion
extracted from high-energy ARPES data we first calculate the
${\BK}$-resolved spectral function $A({\BK},\omega)$ for
Sr$_2$RuO$_4$ (for details see \cite{kink_paper}) defined by
  \begin{eqnarray}
    A({\BK},\omega)&=&-\frac{1}{\pi}{\rm{Im Tr}} \BG({\BK},\omega)
    \label{aofomega}.
  \end{eqnarray}
This quantity  is determined by the diagonal elements of the Green
function matrix in orbital space
  \begin{eqnarray}
    \BG({\BK},\omega)&=&
    [\omega - \BS(\omega) - \bm{H}^{\text{WF}}_{0}({\BK})]^{-1},
    \label{spec_fun}
  \end{eqnarray}
where $\bm{H}^{\text{WF}}$ is the few-orbital Hamiltonian with
t$_{2g}$ symmetry obtained by WF projection. The corresponding
eigenvalues are pictured in Fig.~\ref{lda_bands3} as black lines.
Since  QMC only provides the self-energy $\Sigma$ for Matsubara
frequencies $i\omega_n$ and the \emph{local} spectrum $A(\omega)$,
the calculation of $A({\BK},\omega)$ requires a method to compute
$\Sigma$ for real frequencies $\omega$. This is achieved by first
employing Kramers-Kronig to obtain
  \begin{eqnarray}    \label{fullGF}
    G(\omega+i\eta) =
    \int\limits_{-\infty}^{\infty}d\omega' \;\frac{
      A(\omega')}{\omega - \omega' + i\eta}.
  \end{eqnarray}
The local Green function and the
complex self-energy are related by the ${\BK}$-integrated Dyson equation
  \begin{eqnarray}
    \bm{G}(\omega)=\int\limits_{\text{BZ}}\!d{\BK}\;[\omega+\mu -
    \bm{\Sigma}(\omega) - \bm{h}^{\text{eff}}_{0}({\BK})]^{-1}.
    \label{Ham_intg}
  \end{eqnarray}
Solving Eq.~(\ref{Ham_intg}) for given $G(\omega)$ for the
self-energy $\Sigma(\omega)$ leads to the results presented in Fig.~\ref{sigma}, which
are then inserted into Eq.~(\ref{spec_fun}) to obtain the spectral function
$A({\bf k},\omega)$ from Eq.~(\ref{aofomega}). The maxima of $A({\vec k}, \omega)$ obtained
from this procedure are shown in Fig.~\ref{dispers_comp} as black dots
without error bars. Compared to experimental results our theoretical quasiparticle bands
are shifted by -0.08 eV.

ARPES data directly provide the energy distribution
curves,~\cite{Suga_ARPES, Suga_JESRP} whose second derivative
represents the dispersion, which is shown in Fig.~\ref{dispers_comp}
by a ``rainbow scale'', red being the highest energy. Closed circles
with error bars denote the peak  positions of the second derivative.
The Fermi momenta k$_{\textrm{F}}$ estimated from the momentum
distribution curves (MDCs)~\cite{Suga_ARPES} at E$_{\textrm{F}}$ are
shown by triangles. We see that in the experiment there are
two bands crossing the Fermi level at different {\bf k}-points in
the (0,0)-($\pi$,0) direction. There are also two bands in the
calculated dispersion curve but they are almost degenerate.
Nevertheless the position of the bottom of quasiparticle bands, the
intersection with the Fermi level, and the shape of the experimental
and theoretical dispersions qualitatively agree.

\section{Conclusion}
\label{conclusion}
In this paper we focused on the long-standing controversy concerning
the strength of correlation effects in Sr$_2$RuO$_4$, i.e., on the
question whether Sr$_2$RuO$_4$ should be considered a strongly
correlated electron material. This is generally the case if the
ratio of Coulomb interaction and kinetic energy (bandwidth) is
larger than unity. In particular, electronic correlations lead to a
typical redistribution of spectral weight and thereby to the
formation of well pronounced lower and upper Hubbard bands (LHB,
UHB). We note that even on the experimental level the unambiguous
identification of a maximum in the spectrum of Sr$_2$RuO$_4$ is made
complicated by the overlap of the Ru-4$d$ and O-2$p$ bands.

To answer this question we first calculated the electronic structure
of Sr$_2$RuO$_4$ within the conventional band theory using LDA.
The correlations were then taken into account in the framework of
the LDA+DMFT(QMC) scheme using {\it ab initio} values for the
Coulomb and Hund exchange parameters. We found that the ratio of
Coulomb interaction and bandwidth is indeed larger than unity in
Sr$_2$RuO$_4$, despite the rather extended $4d$-states of Ru,
leading to a distinctive redistribution of spectral weight and to
the formation of a well-pronounced LHB. By comparing our theoretical
spectra with XPS experiments we unambiguously identified this LHB
with the structure observed at -3 eV. By contrast, the LDA DOS shows
no distinctive feature at that energy.

To describe the experimental spectra in a wide energy range we
employed a Wannier function formalism to transform the self-energy
operator back to the full-orbital basis. The theoretical spectra
obtained in that way agree very well with high-photon energy
photoemission data; in particular, they reproduce the shoulder in
the spectrum caused by the LHB. The basic features of the low photon
energy UPS and intermediate energy PES spectra are also reproduced
by the LDA+DMFT(QMC). Quasiparticle bands induced by correlations
with mass renormalization of about 2.5 agree well with results from
ARPES, dHvA and infrared optical experiments. The LDA+DMFT(QMC)
derived quasiparticle bands are even in quantitative agreement with
the dispersion extracted from ARPES data. Taken together these
results provide clear evidence for strong electronic correlation
effects in Sr$_2$RuO$_4$. Hence, although Sr$_2$RuO$_4$ is  a $4d$
system it must be regarded as a strongly correlated electron
material.

\section{Acknowledgment}

We thank R. Claessen, E.Z. Kurmaev, and A. Lichtenstein for helpful
discussions. This work was supported by Russian Basic Research
Foundation grants RFFI-GFEN-03-02-39024, RFFI-04-02-16096,
RFFI-05-02-17244, RFFI-05-02-16301, by the Deutsche
Forschungsgemeinschaft through Sonderforschungsbereiche 484 and 602,
and in part by programs of the Presidium of the Russian Academy of
Sciences (RAS) ``Quantum macrophysics'' and the Division of Physical
Sciences of the RAS ``Strongly correlated electrons in
semiconductors, metals, superconductors and magnetic materials''. IN
and ZP acknowledge support by the Russian Science Support
Foundation, Dynasty Foundation and International Center for
Fundamental Physics in Moscow. IN also appreciates the support from
the Grant of President of Russian Federation for young scientists
MK-2118.2005.02.

\appendix*

\section{Analysis of LDA+DMFT(QMC) spectrum}
\label{thomas}
An important first step in interpreting the structures in the DOS in
Fig.~\ref{qmc_dos} is to identify purely atomic excitations. These will of
course be shifted and broadened by correlation effects, but should
nevertheless still be prominent. These excitations can readily be
obtained from the atomic level picture. We denote the bare level
energy of the $xz$, $yz$ orbitals as
$\epsilon_{xz,yz}\equiv\epsilon_0$. Note that we do not know this
value {\it a priori}, since it is neither the center-of-gravity of the bare LDA-DOS
nor a particular peak position in the correlated DOS. It is, in fact,
the unknown shift due to the double-counting correction of the LDA.

Hund exchange and Coulomb parameters from constrained LDA are
$J=0.7$ eV and $\bar U=1.7$ eV. For three t$_{2g}$-orbitals we have
$U'=\bar U=1.7$ eV and thus $U=U'+2J=3.1$ eV. Finally,
the splitting of the centers of gravity of the $xy$ and $xz$, $yz$ DOS
is $\Delta\epsilon=\epsilon_{xy}-\epsilon_{xz,yz}=0.1$ eV.

With these information we can construct the basis spanning the ground
state and then calculate the excited states and their relative
energies. Concentrating on a particular basis vector of the ground
state manifold, we obtain the scheme in Tab.~\ref{tab1}, where
\begin{table*}[htb]
\begin{center}
\caption[]{Ground-state as well as single-particle excitations
  and their energies.\label{tab1}}
\begin{ruledtabular}
\begin{tabular}{clllccc}
&$|xz\rangle$
&$|yz\rangle$
&$|xy\rangle$
&Energy $E_\alpha$
&
\begin{minipage}[t]{4cm}
Exitation
energy $\epsilon$
\\[-6mm]\mbox{}\end{minipage}
&value, eV\\
\hline
Gs
&$|\uparrow \downarrow\rangle$
&$|\uparrow\rangle$
&$|\uparrow\rangle$
&4$\epsilon_0+\Delta\epsilon+6U-13J$&\\
&\multicolumn{5}{c}{\begin{minipage}[t]{5cm}occupied states\\$\epsilon=E_{\rm
    GS}-E_\alpha$\\[-2mm]\mbox{}
\end{minipage}}\\
c$_{xy}^{\uparrow}$
&$|\uparrow \downarrow\rangle$
&$|\uparrow\rangle$
&$|0\rangle$
&3$\epsilon_0+3U-5J$
&$\epsilon_0+\Delta\epsilon+3U-8J$=$\epsilon_0+3.8$
&-3.2\\
c$_{xz}^{\uparrow}$
&$|\downarrow\rangle$
&$|\uparrow\rangle$
&$|\uparrow\rangle$
&3$\epsilon_0+\Delta\epsilon+3U-7J$
&$\epsilon_0+3U-6J$=$\epsilon_0+5.1$
&-1.9\\
c$_{yz}^{\uparrow}$
&$|\uparrow\downarrow\rangle$
&$|0\rangle$
&$|\uparrow\rangle$
&3$\epsilon_0+\Delta\epsilon+3U-5J$
&$\epsilon_0+3U-8J$=$\epsilon_0+3.7$
&-3.3\\
c$_{xz}^{\downarrow}$
&$|\uparrow\rangle$
&$|\uparrow\rangle$
&$|\uparrow\rangle$
&3$\epsilon_0+\Delta\epsilon+3U-9J$
&$\epsilon_0+3U-4J$=$\epsilon_0+6.5$
&-0.5\\
&\multicolumn{5}{c}{\begin{minipage}[t]{5cm}empty states\\
$\epsilon=E_\alpha-E_{\rm GS}$\\[-2mm]\mbox{}
\end{minipage}}\\
c$_{xy}^{\dag}$
&$|\uparrow\downarrow\rangle$
&$|\uparrow\rangle$
&$|\uparrow\downarrow\rangle$
&5$\epsilon_0+2\Delta\epsilon+10U-20J$
&$\epsilon_0+\Delta\epsilon+4U-7J$=$\epsilon_0+7.5$
&0.5\\
c$_{yz}^{\dag}$
&$|\uparrow\downarrow\rangle$
&$|\uparrow\downarrow\rangle$
&$|\uparrow\rangle$
&5$\epsilon_0+\Delta\epsilon+10U-20J$
&$\epsilon_0+4U-7J$=$\epsilon_0+7.4$
&0.4
\end{tabular}
\end{ruledtabular}
\end{center}
\end{table*}
we show the states contributing to the possible single-particle
excitations. The corresponding excitation energies with the unknown
level shift $\epsilon_0$ are listed in the third column. Obviously,
the transition c$_{xy}^{\uparrow}|Gs\rangle$ represents the
excitation with the lowest energy of the $xy$-DOS, i.e., should be
identified with the position of lowest peak at energy E$_1=-3.2$ eV
in the LDA+DMFT(QMC) DOS in Fig.~\ref{qmc_dos}, leading to
$\epsilon_0=-7$ eV and the final numerical values in the last column
of Tab.~\ref{tab1}.

The identification of the further structures is now
straightforward. For the $xy$ orbital, we must attribute the peak
around $E=0.5$ eV to the excited states
$c_{xy}^\dag|GS\rangle$. Likewise, in the $xz$, $yz$ manifold the peak
at $E\approx-3.5$ eV corresponds to $c_{yz}^\uparrow|GS\rangle$, and
the broad structure at $E\approx-1.3$ eV to a superposition of
$c_{xz}^\uparrow|GS\rangle$ and $c_{xz}^\downarrow|GS\rangle$.
The peak at $E\approx0$ eV finally can be identified with
$c_{yz}^\dag|GS\rangle$.

Note that the peak in the $xy$-DOS at $E\approx-0.8$ eV has no
correspondence in the atomic scheme. It could, however, be due to a
finite admixture of the state
$|\uparrow\rangle|\uparrow\rangle|\uparrow\downarrow\rangle$ to the
ground-state of the interacting system. This state would then allow
for an excitation involving $c_{xy}^\downarrow$ with an excitation
energy $-0.6$ eV.

\end{document}